\documentclass[11pt,a4paper]{article}

\usepackage[margin=1in]{geometry}
\usepackage{amsmath,amssymb,amsthm}
\usepackage{booktabs}
\usepackage{array}
\usepackage{multirow}
\usepackage{graphicx}
\usepackage{xcolor}
\usepackage{hyperref}
\usepackage[ruled,vlined ,linesnumbered]{algorithm2e}
\usepackage{enumitem}
\usepackage{float}
\usepackage{tabularx}
\usepackage{colortbl}
\usepackage{natbib}
\usepackage{subcaption}
\usepackage{authblk}
\hypersetup{
    colorlinks=true,
    linkcolor=blue!70!black,
    citecolor=blue!70!black,
    urlcolor=blue!70!black,
}

\newtheorem{definition}{Definition}

\newcommand{\tlag}{\textsc{T-LAG}}

\title{PulseBench-Tab: A Multilingual Benchmark for Table Extraction\\with Graph-Based Evaluation}

\author{%
  Ritvik Pandey\textsuperscript{1} \quad
  Sid Manchkanti\textsuperscript{1} \quad
  Mohammed Wazir Adain\textsuperscript{2} \\
  Mohammed Hadi\textsuperscript{3} \quad
  Dushyanth Sekhar\textsuperscript{3}
  \vspace{0.5em} \\
  {\small \textsuperscript{1}Pulse AI \quad
  \textsuperscript{2}Georgia Institute of Technology \quad
  \textsuperscript{3}S\&P Global, Enterprise Data Organization}
  \vspace{0.3em} \\
  {\small Email: \texttt{\{ritvik, sid\}@runpulse.com}}
}\date{April 2026}

\begin{document}
\maketitle

\begin{abstract}
We introduce PulseBench-Tab, an open multilingual benchmark for evaluating table extraction from document images. The benchmark comprises 1,820 human-annotated tables spanning 9 languages and 4 scripts (Latin, CJK, Arabic, Cyrillic), drawn from 380 real-world source documents including financial filings, government reports, and regulatory disclosures. Tables range from 2 to 1,183 cells, with 48.1\% containing merged or spanning cells. Alongside the dataset, we propose \tlag{} (Table Logical Adjacency Graph), a novel evaluation metric that models tables as directed graphs over cell adjacencies and computes structural and content fidelity in a single score via optimal bipartite matching. We evaluate 9 commercial and open-source table extraction systems across the benchmark and report per-language breakdowns. The full dataset, scoring code, and all provider outputs are publicly available.
\end{abstract}

\section{Introduction}\label{sec:intro}

Table extraction from document images is a core task in document intelligence, underpinning downstream applications in financial analysis, regulatory compliance, and enterprise data processing. A table is a two-dimensional structure where position is meaning: a number in the wrong cell is not merely an OCR error but a semantic failure that propagates silently through downstream pipelines~\citep{gartlehner2025promise}.

Despite the importance of the task, existing benchmarks and metrics have well-documented limitations. TEDS~\citep{zhong2020image} operates on DOM trees and conflates formatting conventions (e.g., \texttt{<thead>} wrappers) with structural errors. Sequence-alignment approaches~\citep{reducto2024rdtablebench} flatten tables into one-dimensional sequences, losing the distinction between horizontal and vertical adjacency. Most benchmarks evaluate primarily on English documents, leaving performance on non-Latin scripts, right-to-left languages, and CJK text largely unmeasured.

PulseBench-Tab addresses these gaps along three axes:

\begin{enumerate}[nosep]
    \item \textbf{Multilingual coverage.} 9 languages across 4 scripts, with intentional representation of Arabic (RTL), Chinese, Japanese, and Korean.
    \item \textbf{Structural complexity.} Tables range from simple grids to dense spreadsheets with multi-level headers, merged cells, and over 1,000 cells.
    \item \textbf{2D-aware evaluation.} \tlag{} evaluates tables as directed graphs, preserving the distinction between horizontal and vertical adjacency, and uses optimal bipartite matching rather than greedy or sequence-based alignment.
\end{enumerate}

The dataset, scoring code, and all provider outputs are released publicly.\footnote{Dataset: \url{https://huggingface.co/datasets/pulse-ai/PulseBench-Tab}} \footnote{Code: \url{https://github.com/Pulse-Software-Corp/PulseBench-Tab}}

\section{Related Work}\label{sec:related}

\paragraph{Benchmarks.}
OmniDocBench~\citep{ouyang2025omnidocbench} evaluates tables alongside text and formulas using TEDS, with attribute-level granularity across table types. However, TEDS conflates formatting with structure, and the benchmark covers only English and Chinese. RD-TableBench~\citep{reducto2024rdtablebench} provides 1,000 publicly available table images across 15 languages with Needleman-Wunsch scoring, but the linearization of tables into sequences loses 2D structural information. SCORE-Bench~\citep{unstructured2025scorebench} separates content accuracy from index accuracy and introduces spatial tolerance for structural ambiguity, but does not provide per-language or per-complexity breakdowns. PubTabNet~\citep{zhong2020image} and FinTabNet~\citep{zheng2021global} are large-scale datasets but focus on English scientific and financial tables respectively.

\paragraph{Metrics.}
TEDS~\citep{zhong2020image} computes tree edit distance on the HTML DOM and normalizes by tree size. GriTS~\citep{smock2023grits} introduces grid-based matching but uses greedy alignment rather than optimal matching and does not distinguish edge directions. DAR (Digit Accuracy Rate) counts cell-level exact matches but is insensitive to positional errors. \tlag{} differs from all of these by operating on directed adjacency edges with optimal bipartite matching, capturing both content and structure in a single pass.

\section{Dataset}\label{sec:dataset}

\subsection{Overview}

PulseBench-Tab contains 1,820 table images extracted from 380 unique source documents. Each image depicts a single table, and its ground truth is a human-annotated HTML table with full structural markup (\texttt{rowspan}, \texttt{colspan}, \texttt{<th>}, \texttt{<td>}).

Source documents include financial filings, government reports, corporate disclosures, and regulatory filings across all 9 target languages. Tables range from simple 2-cell headers to dense spreadsheets with 1,183 cells. 48.1\% of tables contain merged or spanning cells.

\subsection{Language Distribution}

The dataset spans 9 languages across 4 scripts (Latin, CJK, Arabic, Cyrillic):

\begin{table}[H]
\centering
\begin{tabular}{lrr}
    \toprule
    \textbf{Language} & \textbf{Samples} & \textbf{\% of Dataset} \\
    \midrule
    English  & 594 & 32.6\% \\
    Chinese  & 213 & 11.7\% \\
    Spanish  & 176 & 9.7\%  \\
    Russian  & 170 & 9.3\%  \\
    French   & 165 & 9.1\%  \\
    Japanese & 159 & 8.7\%  \\
    Arabic   & 146 & 8.0\%  \\
    German   & 113 & 6.2\%  \\
    Korean   & 84  & 4.6\%  \\
    \midrule
    \textbf{Total} & \textbf{1,820} & \textbf{100\%} \\
    \bottomrule
\end{tabular}
\caption{Language distribution in PulseBench-Tab.}
\label{tab:language-dist}
\end{table}

\begin{figure}[H]
\centering
\includegraphics[width=0.9\textwidth]{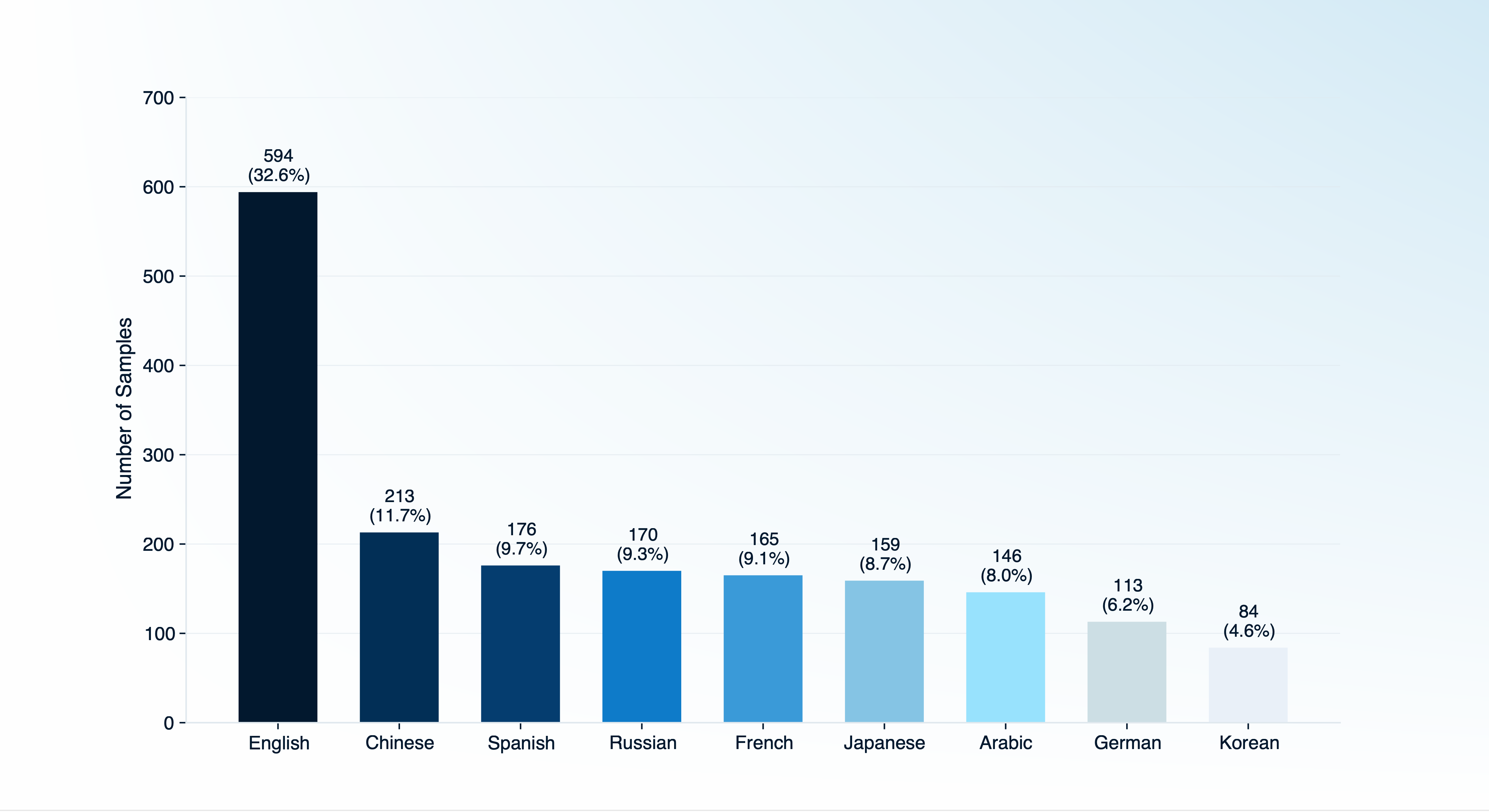}
\caption{Language distribution across PulseBench-Tab. English comprises 32.6\% of samples; the remaining 8 languages provide coverage across Latin, CJK, Arabic, and Cyrillic scripts.}
\label{fig:lang-dist}
\end{figure}

\subsection{Table Complexity}

\begin{table}[H]
\centering
\begin{tabular}{lrrrrrr}
    \toprule
    \textbf{Metric} & \textbf{Mean} & \textbf{Min} & \textbf{Max} & \textbf{P25} & \textbf{P75} & \textbf{P90} \\
    \midrule
    Rows           & 11.3 & 1 & 65    & 5  & 14 & 24  \\
    Columns        & 5.0  & 2 & 28    & 3  & 6  & 8   \\
    Cells          & 54.1 & 2 & 1,183 & 18 & 65 & 112 \\
    Spanning cells & 1.9  & 0 & 38    & 0  & 3  & 6   \\
    \bottomrule
\end{tabular}
\caption{Table complexity statistics across the 1,820 samples.}
\label{tab:complexity}
\end{table}

Tables span a wide range of difficulty. 27.5\% have 20 cells or fewer; 4.0\% exceed 200 cells. The distribution of cell counts is shown in Table~\ref{tab:size-dist}.

\begin{table}[H]
\centering
\begin{tabular}{lrr}
    \toprule
    \textbf{Cell Count} & \textbf{Samples} & \textbf{\% of Dataset} \\
    \midrule
    1--20   & 500 & 27.5\% \\
    21--50  & 647 & 35.5\% \\
    51--100 & 397 & 21.8\% \\
    101--200 & 204 & 11.2\% \\
    201+    & 72  & 4.0\%  \\
    \bottomrule
\end{tabular}
\caption{Distribution of table sizes by cell count.}
\label{tab:size-dist}
\end{table}

\begin{figure}[H]
\centering
\includegraphics[width=0.7\textwidth]{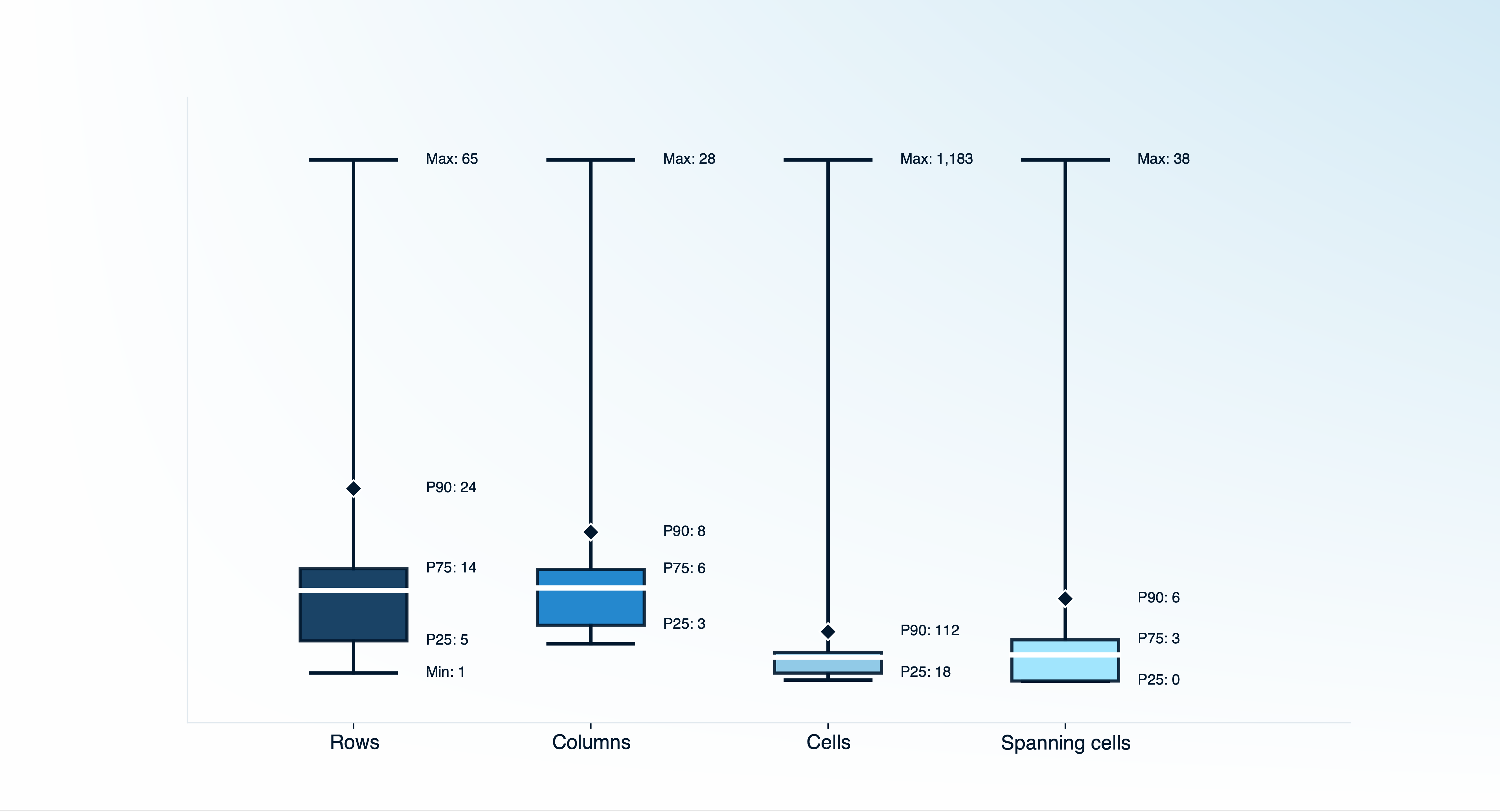}
\caption{Distribution of table complexity. Tables range from 2 to 1,183 cells, with 48.1\% containing merged or spanning cells. The long tail of large tables stresses both OCR and structural recovery.}
\label{fig:complexity}
\end{figure}

\subsection{Ground Truth Construction}

Ground truth was produced through a multi-stage annotation pipeline involving human annotators across 8 labeling rounds, with dedicated review teams for each target language and script system.

\begin{enumerate}[nosep]
    \item \textbf{Expert annotation.} A distributed team of domain specialists - including native speakers for each of the 9 target languages - manually annotated table structure and content in HTML, producing full structural markup with \texttt{rowspan}/\texttt{colspan} attributes. Annotators were trained on a 40-page annotation guideline covering spanning cells, nested headers, and script-specific edge cases (e.g., right-to-left column ordering in Arabic, full-width punctuation in CJK).
    \item \textbf{Cross-lingual quality control.} All candidate samples underwent independent cross-review by language-specialist reviewers. 
    \item \textbf{Adversarial review.} Every ground truth table was independently re-examined against its source image in a cell-by-cell audit, flagging structural mismatches (wrong row/column counts, missing spans) and content errors (character-level misreads, shifted data). Each proposed correction required independent confirmation before acceptance. 
\end{enumerate}

The final dataset of 1,820 samples represents the subset that passed all quality gates. 

\section{Evaluation Methodology: \tlag{}}\label{sec:tlag}

\tlag{} (Table Logical Adjacency Graph) evaluates table extraction quality by modeling tables as directed graphs and computing a weighted $F_1$ score over optimally matched edges. The metric captures both structural fidelity (are cells in the right positions?) and content accuracy (is the text correct?) in a single number.

\subsection{Why Edges, Not Nodes}

A natural question is why \tlag{} operates on \textit{edges} (adjacency relationships) rather than \textit{nodes} (individual cells). The reason is that edges capture both the content of cells and the structural relationships between them. A node-based metric can verify that a cell's text is correct but cannot verify that the cell is in the correct position relative to its neighbors. Two tables with identical cell contents but different layouts (e.g., transposed columns) would score identically under a node-based metric but differ under an edge-based one, because the adjacency relationships change. Edges are the minimal unit that jointly encodes content and structure.

\subsection{Pipeline Overview}

The \tlag{} pipeline has five phases:

\[
\boxed{\text{Parse}} \;\longrightarrow\;
\boxed{\text{Extract}} \;\longrightarrow\;
\boxed{\text{Weight}} \;\longrightarrow\;
\boxed{\text{Match}} \;\longrightarrow\;
\boxed{\text{Score}}
\]

\subsubsection{Phase 1: HTML to Grid Cell Matrix}

Both the ground truth and predicted HTML tables are parsed into a cell-position grid matrix $M[R \times C]$, preserving all \texttt{rowspan} and \texttt{colspan} attributes. Each cell is assigned a unique integer ID. A cell with \texttt{rowspan}=$s_r$ and \texttt{colspan}=$s_c$ at logical position $(r, c)$ occupies all grid positions $(r+i, c+j)$ for $i \in [0, s_r)$, $j \in [0, s_c)$, sharing the same ID.

\begin{equation}
    M[r, c] = \text{cell\_id} \quad \text{for } 0 \le r < R,\; 0 \le c < C
\end{equation}

No column reversal is applied for right-to-left tables. The HTML \texttt{dir="rtl"} attribute is a visual rendering hint that does not change the logical DOM order of \texttt{<td>} elements.

\subsubsection{Phase 2: Extract Directed Edges}

We walk the grid to emit directed adjacency edges:

\begin{definition}[RIGHT edge]
For every grid position $(r, c)$ where $c + 1 < C$, if $M[r, c]$ and $M[r, c+1]$ have different cell IDs, emit:
\[
    e = \bigl(M[r,c] \xrightarrow{\text{RIGHT}} M[r, c+1]\bigr)
\]
\end{definition}

\begin{definition}[BELOW edge]
For every grid position $(r, c)$ where $r + 1 < R$, if $M[r, c]$ and $M[r+1, c]$ have different cell IDs, emit:
\[
    e = \bigl(M[r,c] \xrightarrow{\text{BELOW}} M[r+1, c]\bigr)
\]
\end{definition}

The cell ID condition suppresses edges within spanning cells: a cell with \texttt{colspan=2} occupying $(r, 0)$ and $(r, 1)$ does not generate a RIGHT edge between those positions. Edges are deduplicated by the tuple $(\text{source\_id}, \text{target\_id}, \text{direction})$ to prevent large merged cells from dominating the score.

\subsubsection{Phase 3: Edge Weighting via the $\Psi$ Function}

For every candidate pair of ground truth edge $e_{gt}$ and predicted edge $e_{pred}$ sharing the same direction, we compute a similarity weight:

\begin{equation}\label{eq:edge-weight}
    w(e_{gt}, e_{pred}) = \Psi(\text{src}_{gt},\; \text{src}_{pred}) \;\times\; \Psi(\text{tgt}_{gt},\; \text{tgt}_{pred})
\end{equation}

The $\Psi$ function is a text similarity kernel. Both texts are first normalized: leading/trailing whitespace is stripped, Unicode dashes are normalized, internal whitespace is collapsed, and null markers ($\varepsilon$, ``-'', ``--'', ``---'', ``...'', ``n/a'', ``na'', ``none'', ``nil'') are mapped to a sentinel value \texttt{[NULL]}.

\begin{definition}[$\Psi$ -- Text Similarity Kernel]
Given normalized texts $g$ and $p$:
\begin{equation}\label{eq:psi}
\Psi(g, p) = \begin{cases}
    1.0 & \text{if } g = \texttt{[NULL]} \text{ and } p = \texttt{[NULL]} \\[4pt]
    0.0 & \text{if exactly one of } g, p = \texttt{[NULL]} \\[4pt]
    \displaystyle\left(1 - \frac{d_{\text{Lev}}(g,\, p)}{\max(|g|, |p|)}\right)^{\!7} & \text{otherwise}
\end{cases}
\end{equation}
where $d_{\text{Lev}}$ is the Levenshtein edit distance and $|\cdot|$ denotes string length.
\end{definition}

All comparison is case-sensitive with no numeric normalization. The exponent $k=7$ creates aggressive decay: a raw similarity of 0.90 yields $\Psi = 0.478$, and 0.80 yields $\Psi = 0.210$. This sharpness is deliberate; in production table extraction, a number in a neighboring cell is not ``almost right.''

\begin{table}[H]
\centering
\begin{tabular}{rrrrr}
    \toprule
    Raw similarity & Interpretation & $k=3$ & $k=5$ & $k=7$ \\
    \midrule
    1.00 & Perfect match & 1.000 & 1.000 & 1.000 \\
    0.95 & 1-char typo in 20-char string & 0.857 & 0.774 & 0.698 \\
    0.90 & Small OCR error & 0.729 & 0.590 & 0.478 \\
    0.80 & Significant Error & 0.512 & 0.328 & 0.210 \\
    0.70 & Major Divergence & 0.343 & 0.168 & 0.082 \\
    0.50 & Half Wrong & 0.125 & 0.031 & 0.008 \\
    \bottomrule
\end{tabular}
\caption{Effect of power-decay exponent on the $\Psi$ function. At $k=7$, even small text differences are heavily penalized.}
\label{tab:decay}
\end{table}

\subsubsection{Phase 4: Hungarian Matching}

We construct a weight matrix $W$ between all ground truth edges and all predicted edges, where $W[i,j] = w(e^{gt}_i, e^{pred}_j)$ if the edges share the same direction, and $W[i,j] = 0$ otherwise. The Hungarian algorithm~\citep{kuhn1955hungarian} finds the optimal one-to-one assignment maximizing total matched weight.

This matching has three key properties:
\begin{itemize}[nosep]
    \item \textbf{Optimal}: globally optimal assignment, not greedy.
    \item \textbf{One-to-one}: no double counting of edges.
    \item \textbf{Direction-aware}: RIGHT edges only match RIGHT; BELOW only matches BELOW.
\end{itemize}

\subsubsection{Phase 5: Score}

Let $S$ denote the total matched weight from the Hungarian assignment, $|\mathcal{E}_{gt}|$ the number of ground truth edges, and $|\mathcal{E}_{pred}|$ the number of predicted edges. Then:

\begin{align}
    \text{Precision} &= \frac{S}{|\mathcal{E}_{pred}|} \label{eq:precision} \\[4pt]
    \text{Recall}    &= \frac{S}{|\mathcal{E}_{gt}|}   \label{eq:recall} \\[4pt]
    \text{\tlag{}}   &= F_1 = \frac{2 \times \text{Precision} \times \text{Recall}}{\text{Precision} + \text{Recall}} \label{eq:f1}
\end{align}

\paragraph{Interpretation.} Precision measures how many of the predicted edges correspond to actual edges in the ground truth, weighted by text similarity. A system that hallucinates extra rows or columns will have low precision due to unmatched predicted edges. Recall measures how many of the ground truth edges were recovered by the prediction. A system that drops rows, misses merged cells, or truncates tables will have low recall due to unmatched ground truth edges. The $F_1$ balances both failure modes.

\paragraph{Single-cell fallback.} When both tables have zero edges (single-cell tables), the score falls back to $\Psi$ applied directly to the cell text, preventing degenerate 1.0 scores on mismatched single-cell tables.

\subsection{Exponent Selection ($k = 7$)}\label{sec:k-selection}

The $\Psi$ exponent $k$ controls how aggressively partial text matches are penalized. We set $k = 7$ based on the fidelity requirements of enterprise document processing. In financial filings, regulatory disclosures, and compliance reporting, a cell with the wrong number is not ``almost correct'' - it is unusable. A single misread digit in a revenue figure or regulatory threshold can invalidate an entire downstream analysis.

At $k = 7$, a cell that is 90\% character-accurate contributes $\Psi = 0.478$, and an 80\%-accurate cell contributes $\Psi = 0.210$. This steep decay reflects the operational reality that near-miss extraction is not meaningfully better than outright failure for production use cases. A table with approximate numbers cannot be used for computation, reporting, or compliance verification.

\paragraph{Sensitivity analysis.} To confirm that evaluation results are robust to the choice of $k$, we recomputed all scores at $k \in \{8, 9, 11\}$. As shown in Table~\ref{tab:k-sweep}, provider rankings are completely unchanged under stricter settings - the relative ordering is identical at every exponent tested. Absolute scores decrease monotonically as expected, but the magnitude of change from $k = 7$ to $k = 11$ is small (0.5--1.8 points depending on provider), confirming that $k = 7$ already operates in the regime where the metric has converged and further strictness yields diminishing returns.

\begin{table}[H]
\centering
\small
\setlength{\tabcolsep}{5pt}
\begin{tabular}{l|cccc}
    \toprule
    \textbf{Provider} & $k\!=\!7$ & $k\!=\!8$ & $k\!=\!9$ & $k\!=\!11$ \\
    \midrule
    Pulse Ultra 2              & .935 & .933 & .932 & .930 \\
    Gemini 3.1                 & .816 & .812 & .810 & .805 \\
    LlamaParse (Agentic)       & .798 & .793 & .789 & .783 \\
    Reducto (Agentic)          & .795 & .790 & .786 & .778 \\
    Extend                     & .763 & .758 & .753 & .746 \\
    Azure DI                   & .761 & .757 & .752 & .745 \\
    Reducto                    & .718 & .712 & .707 & .700 \\
    AWS Textract               & .603 & .600 & .596 & .591 \\
    Unstructured               & .360 & .355 & .350 & .343 \\
    \bottomrule
\end{tabular}
\caption{Sensitivity of T-LAG scores to the $\Psi$ exponent. Rankings are identical at all tested values of $k$; absolute scores decrease monotonically with diminishing marginal effect.}
\label{tab:k-sweep}
\end{table}

\begin{figure}[H]
\centering
\includegraphics[width=0.85\textwidth]{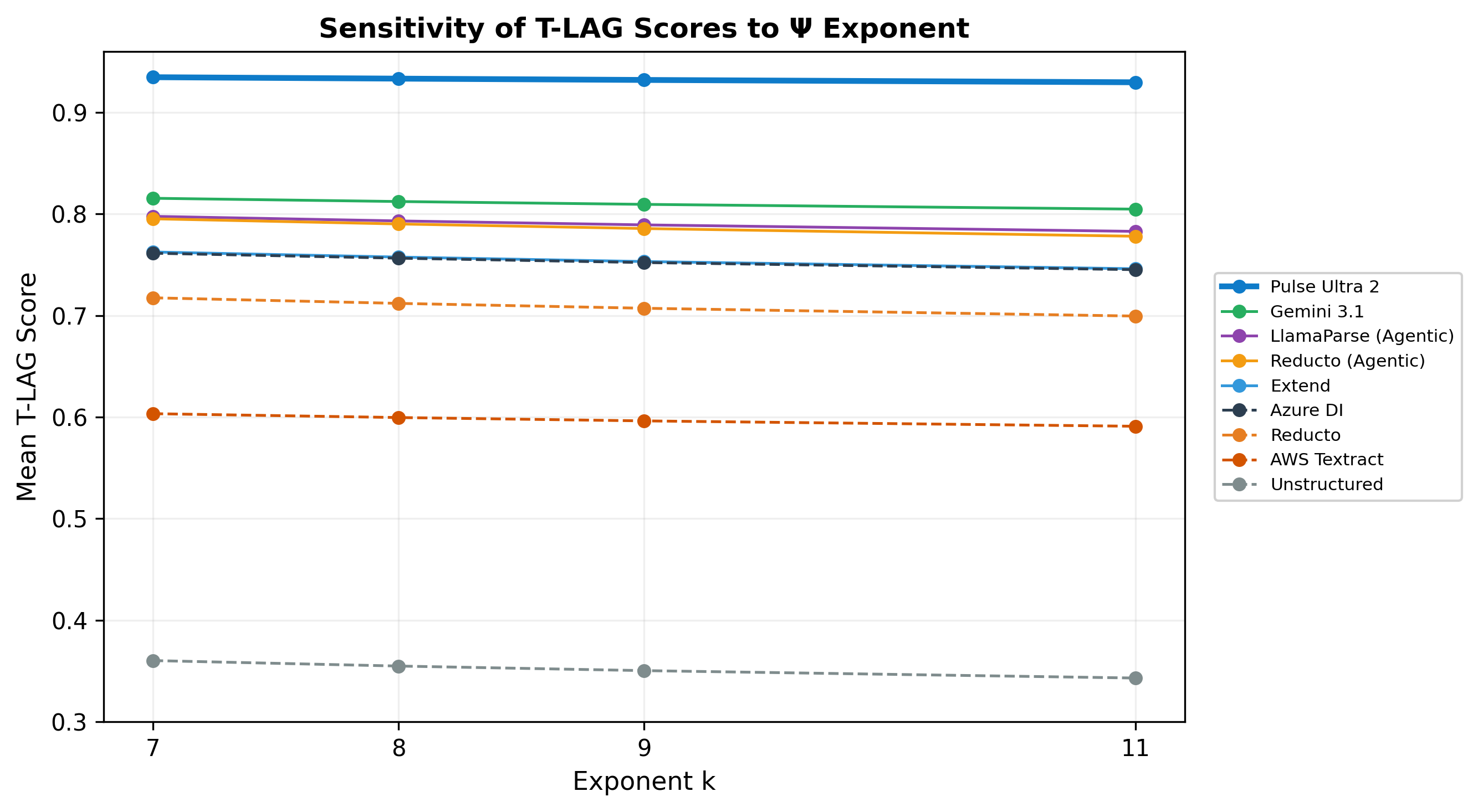}
\caption{Sensitivity of T-LAG scores to the $\Psi$ exponent. All providers decline monotonically as $k$ increases, but rankings remain unchanged. The near-flat curves confirm that $k = 7$ already operates in the converged regime.}
\label{fig:k-sensitivity}
\end{figure}

\section{Experimental Setup}\label{sec:setup}

We evaluated 9 table extraction systems across all 1,820 samples. Systems span commercial document AI APIs, general-purpose large language models, cloud document processing services, and open-source tools. Each provider was called through its publicly available API with default configurations. No system-specific tuning or prompt engineering was applied. Provider outputs are HTML tables, scored against the ground truth using \tlag{}.

To avoid penalizing providers for detection failures (e.g., failing to identify a table region in the image), we adopt an \textit{exclude-missing} scoring mode: providers are scored only on samples where they produced output. Coverage (the percentage of samples each provider returned results for) is reported alongside accuracy.

The evaluated systems are:

\begin{table}[H]
\centering
\begin{tabular}{ll}
    \toprule
    \textbf{Provider} & \textbf{Type} \\
    \midrule
    Pulse Ultra 2                   & Commercial API \\
    Gemini 3.1                      & General-purpose LLM \\
    LlamaParse (Agentic)            & Commercial API \\
    Reducto (Agentic)               & Commercial API \\
    Extend                          & Commercial API \\
    Azure Document Intelligence     & Cloud service \\
    Reducto                         & Commercial API \\
    AWS Textract                    & Cloud service \\
    Unstructured                    & Open-source / API \\
    \bottomrule
\end{tabular}
\caption{Evaluated table extraction systems.}
\label{tab:providers}
\end{table}

\section{Results}\label{sec:results}

\subsection{Overall Rankings}

\begin{table}[H]
\centering
\begin{tabular}{clcccc}
    \toprule
    \textbf{Rank} & \textbf{Provider} & \textbf{T-LAG} & \textbf{Median} & \textbf{Coverage} & \textbf{Perfect} \\
    \midrule
    1 & Pulse Ultra 2              & 0.935 & 1.000 & 100.0\% & 57.9\% \\
    2 & Gemini 3.1                 & 0.816 & 0.939 & 99.5\%  & 28.6\% \\
    3 & LlamaParse (Agentic)       & 0.798 & 0.872 & 94.0\%  & 17.5\% \\
    4 & Reducto (Agentic)          & 0.795 & 0.889 & 78.8\%  & 13.2\% \\
    5 & Extend                     & 0.763 & 0.848 & 91.9\%  & 12.6\% \\
    6 & Azure Document Intelligence & 0.761 & 0.846 & 92.0\%  & 12.6\% \\
    7 & Reducto                    & 0.718 & 0.812 & 80.4\%  & 9.4\%  \\
    8 & AWS Textract               & 0.603 & 0.694 & 98.5\%  & 9.1\%  \\
    9 & Unstructured               & 0.360 & 0.307 & 100.0\% & 2.1\%  \\
    \bottomrule
\end{tabular}
\caption{Overall provider rankings on PulseBench-Tab (1,820 samples, exclude-missing scoring). ``Perfect'' denotes the percentage of samples scoring 1.0.}
\label{tab:rankings}
\end{table}

\begin{figure}[H]
\centering
\includegraphics[width=0.85\textwidth]{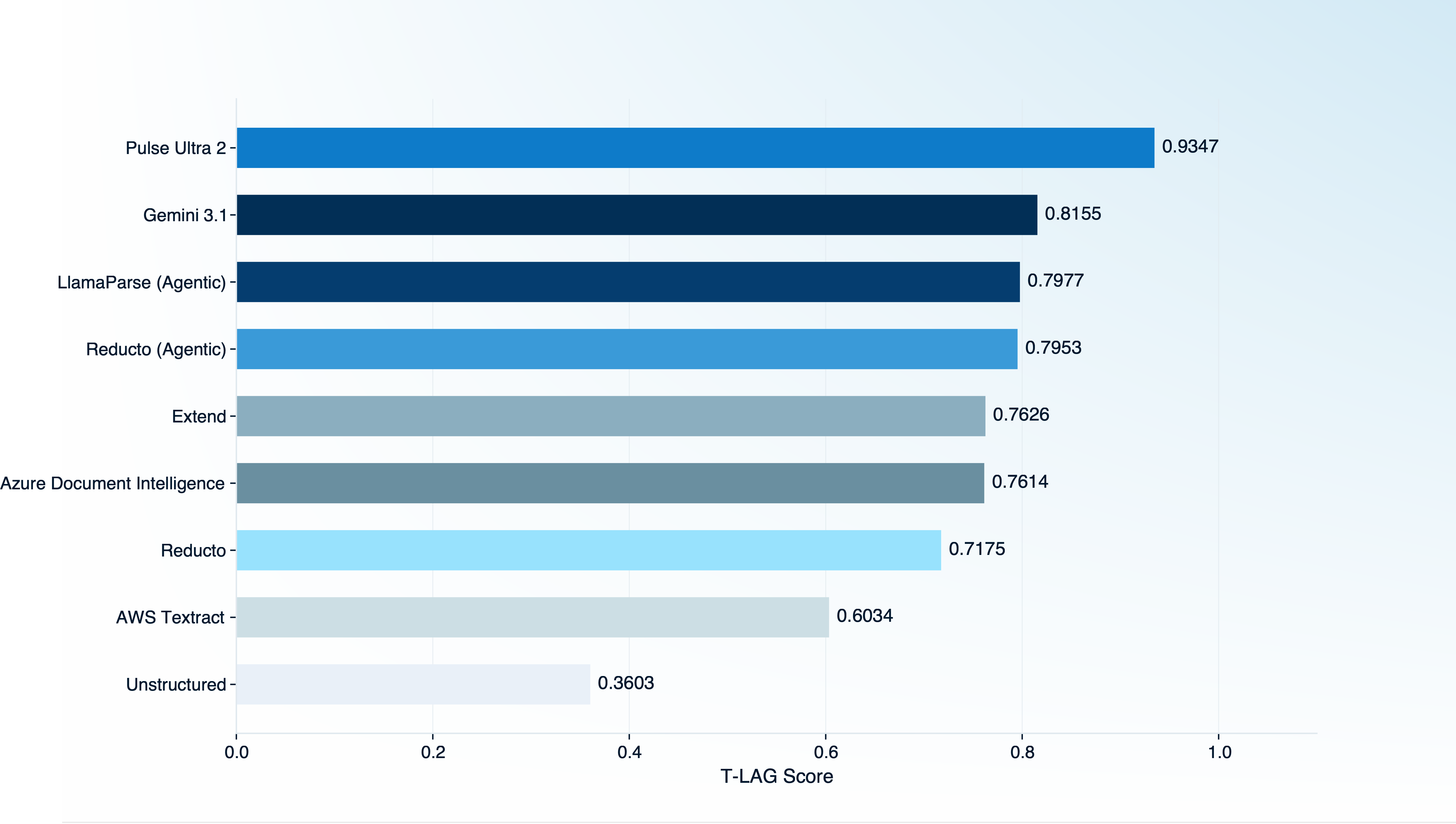}
\caption{Overall T-LAG scores across 9 providers. Pulse Ultra 2 (93.5\%) leads by 11.9 percentage points over the next system. The bottom third of providers score below 72\%, indicating fundamental limitations in multilingual table extraction.}
\label{fig:rankings}
\end{figure}

\subsection{Per-Language Breakdown}

\begin{table}[H]
\centering
\small
\setlength{\tabcolsep}{4pt}
\begin{tabular}{l|ccccccccc}
    \toprule
    \textbf{Language} & \rotatebox{70}{\textbf{Pulse U2}} & \rotatebox{70}{\textbf{Gemini}} & \rotatebox{70}{\textbf{LlamaP.}} & \rotatebox{70}{\textbf{Red. Ag.}} & \rotatebox{70}{\textbf{Extend}} & \rotatebox{70}{\textbf{Azure DI}} & \rotatebox{70}{\textbf{Reducto}} & \rotatebox{70}{\textbf{Textract}} & \rotatebox{70}{\textbf{Unstr.}} \\
    \midrule
    Arabic (146)  & .915 & .660 & .680 & .557 & .605 & .606 & .356 & .240 & .158 \\
    Chinese (213) & .958 & .869 & .812 & .806 & .717 & .714 & .671 & .324 & .173 \\
    English (594) & .910 & .775 & .789 & .774 & .747 & .746 & .717 & .724 & .433 \\
    French (165)  & .972 & .899 & .845 & .858 & .841 & .840 & .804 & .788 & .533 \\
    German (113)  & .953 & .839 & .807 & .831 & .775 & .773 & .786 & .755 & .426 \\
    Japanese (159)& .955 & .827 & .826 & .858 & .802 & .803 & .789 & .479 & .355 \\
    Korean (84)   & .941 & .840 & .804 & .835 & .743 & .742 & .691 & .254 & .137 \\
    Russian (170) & .936 & .870 & .831 & .840 & .799 & .794 & .759 & .640 & .210 \\
    Spanish (176) & .936 & .848 & .797 & .797 & .848 & .848 & .790 & .814 & .563 \\
    \bottomrule
\end{tabular}
\caption{Per-language T-LAG scores. Non-Latin scripts (Arabic, Korean, CJK) show the largest cross-provider variance.}
\label{tab:per-lang}
\end{table}

\begin{figure}[H]
\centering
\includegraphics[width=\textwidth]{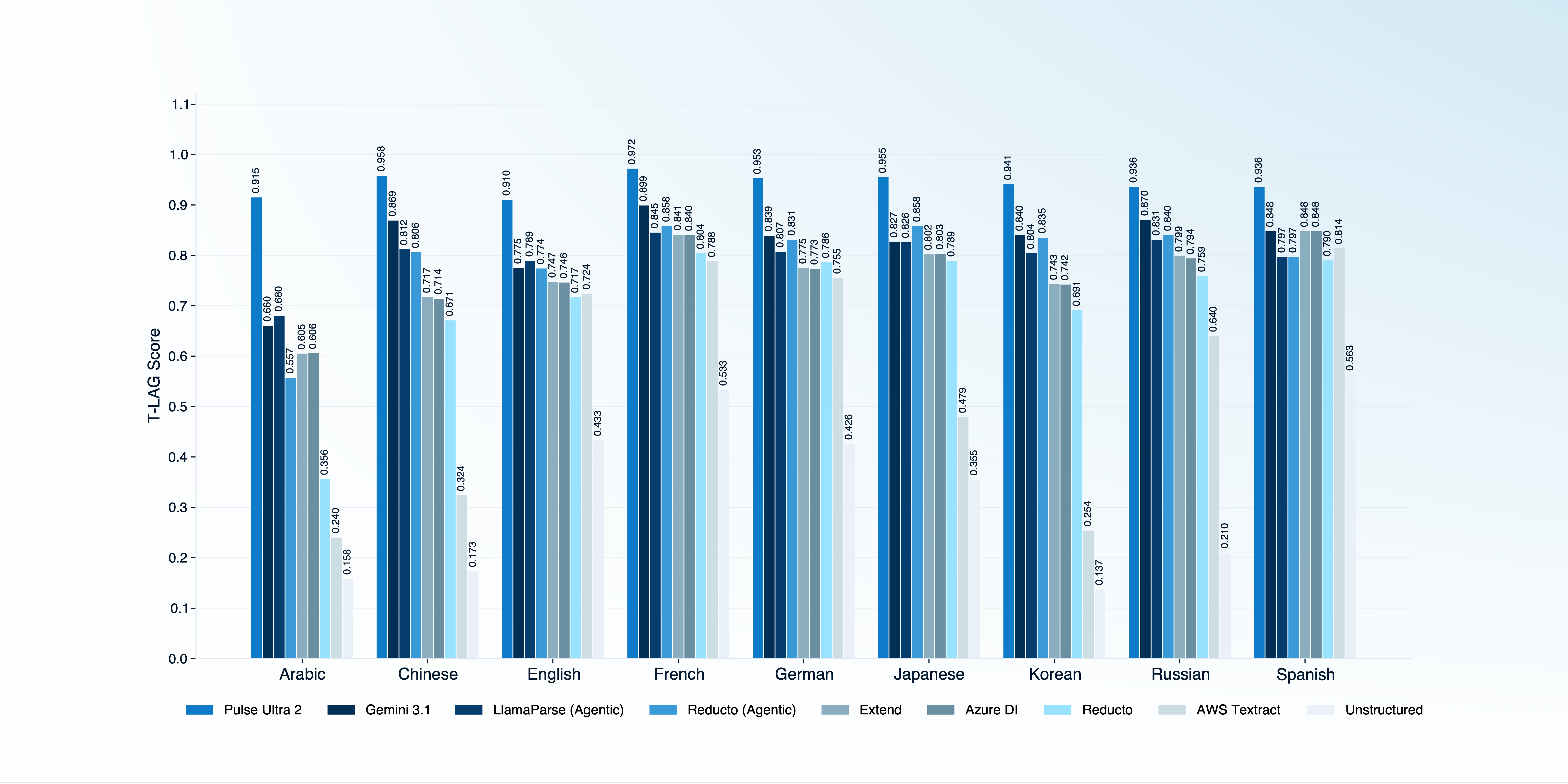}
\caption{Per-language T-LAG scores across providers. Arabic and Korean exhibit the widest cross-provider variance, with top-to-bottom spreads exceeding 75 percentage points.}
\label{fig:lang-heatmap}
\end{figure}

\subsection{Analysis}

Several patterns emerge from the results:

\paragraph{Wide performance gap at the top.} Pulse Ultra 2 is the only provider with a median score of 1.0 (perfect extraction on over half of all tables), leading the benchmark at 93.5\%. The gap to the second-ranked system (Gemini 3.1 at 81.6\%) is 11.9 percentage points - the largest gap anywhere in the ranking. Only one provider exceeds 90\% mean T-LAG.

\paragraph{Structural hallucinations are pervasive.} Even the second-ranked system achieves a perfect-extraction rate of only 28.6\%, meaning it introduces at least one structural or content error on over 70\% of tables. Common failure modes include fabricated rows, invented cell content, incorrect span attributes, and shifted data - errors that are invisible without cell-by-cell verification against the source image. These structural hallucinations make raw LLM or OCR output unreliable for downstream computation without additional verification.

\paragraph{Non-Latin scripts remain challenging.} Arabic and Korean show the largest performance drops across all providers. On Arabic, the gap between the top provider (0.915) and the median provider (0.606) is 30.9 percentage points. AWS Textract and Unstructured score below 0.25 on Arabic and Korean, suggesting fundamental limitations in their OCR or structure recognition pipelines for these scripts.

\paragraph{Coverage varies significantly.} Reducto (Agentic) fails to produce output for 21.2\% of samples, and standard Reducto for 19.6\%. This is relevant for production deployments where silent failures require fallback mechanisms. Coverage is reported alongside accuracy to prevent favorable selection bias.

\paragraph{CJK languages expose OCR weaknesses.} Chinese, Japanese, and Korean tables require character-level precision across large character sets. Providers with strong English performance (e.g., AWS Textract at 0.724 on English) drop sharply on CJK (0.324 on Chinese, 0.254 on Korean).

\paragraph{French and Spanish are easiest.} All providers perform best on French and Spanish, likely due to shared Latin script, well-represented training data, and relatively standard table formatting in government and financial documents from these regions.

\section{Discussion}\label{sec:discussion}

\subsection{Scope}

\paragraph{Domain breadth.} PulseBench-Tab covers a deliberately wide range of document types - financial filings, government reports, regulatory disclosures, historical documents, and scientific tables - across 9 languages and 4 scripts. The dataset includes handwritten annotations, degraded scans from archival sources, and dense multi-page spreadsheets that stress both OCR and structural recovery. That said, certain verticals (medical records, web-native tables, forms with mixed table/non-table regions) remain underrepresented and are natural candidates for future expansion.

\paragraph{Scale} At 1,820 samples the benchmark is larger than most publicly available table extraction benchmarks, but per-language subsets for smaller languages (Korean: 84, German: 113) would benefit from additional samples to narrow confidence intervals. Scaling the dataset further - particularly for underrepresented scripts - is an ongoing effort.

\paragraph{Temporal coverage} The benchmark captures provider performance at a single point in time. As extraction systems improve, periodic re-evaluation will be necessary. The fully open dataset and scoring code are designed to make re-evaluation straightforward.

\paragraph{Scoring philosophy} The $k=7$ exponent in $\Psi$ reflects the fidelity requirements of enterprise document processing, where approximate extraction is operationally equivalent to failure. For use cases with different tolerance thresholds (e.g., exploratory analysis or summarization), the exponent can be adjusted; the scoring code accepts $k$ as a parameter. As shown in Section~\ref{sec:k-selection}, provider rankings are invariant to the choice of $k$.

\subsection{Limitations Future Work}
\paragraph{Single-table assumption} Each image in the benchmark contains exactly one table. Documents with multiple tables per page - or tables embedded within surrounding text - are not evaluated, and provider behavior on multi-table layouts may differ from single-table performance.

\paragraph{Single page table}  Our benchmark doesn't support tables spanning multiple pages yet. All tables are assumed to fit within a single page.

\subsection{Comparison with Existing Metrics}

\tlag{} differs from existing metrics in three principal ways:

\begin{enumerate}[nosep]
    \item \textbf{2D structure preservation.} Unlike TEDS (tree-based) or Needleman-Wunsch (1D sequence), \tlag{} operates on directed edges that explicitly encode horizontal and vertical adjacency as distinct relationships. A column transposition is penalized differently from a content error.
    \item \textbf{Optimal matching.} The Hungarian algorithm provides globally optimal one-to-one assignment, unlike greedy approaches (GriTS) or dynamic programming on linearized sequences.
    \item \textbf{Format agnosticism.} \tlag{} does not penalize HTML formatting conventions. The presence or absence of \texttt{<thead>}, \texttt{<tbody>}, or \texttt{<th>} wrappers does not affect the score, because the metric operates on the parsed grid, not the DOM tree.
\end{enumerate}

\section{Conclusion}\label{sec:conclusion}

We presented PulseBench-Tab, a multilingual benchmark for table extraction comprising 1,820 human-annotated tables across 9 languages, and \tlag{}, a graph-based evaluation metric that captures structural and content fidelity in a single score. Our evaluation of 9 providers reveals significant performance variation across languages, with non-Latin scripts remaining a major challenge for the field. The full dataset, scoring implementation, and all provider outputs are publicly available to support reproducible evaluation and future research.

\section*{Acknowledgements}
PulseBench-Tab was a significant effort across data curation, annotation, validation, and scoring methodology development. We want to thank the people and teams who made it possible.
Thank you to \href{https://www.linkedin.com/in/dushyanth-sekhar-6603105/}{Dushyanth Sekhar}\footnote{\label{fn:spacademic}\textit{Dushyanth Sekhar and Mohammed Hadi are listed as the final two authors to acknowledge their individual academic contributions, specifically methodological guidance on benchmarking approaches and evaluation design. These academic contributions do not constitute a business, commercial, or institutional contribution from S\&P. Author attribution will reference their individual names and company affiliation.}} and \href{https://www.linkedin.com/in/moodyhadi/}{Mohammed Hadi}\textsuperscript{\ref{fn:spacademic}} of S\&P Global's Enterprise Data Organization for their academic contributions\footnote{\textit{Pulse Software, Corp. (``Pulse'') acknowledges and confirms that nothing in the benchmark, its associated materials, or any public-facing content constitutes an endorsement of Pulse by S\&P Global Inc., or any of their affiliates (collectively, ``S\&P''). Likewise, nothing constitutes an endorsement of S\&P by Pulse.}} to the benchmark methodology.
Thank you to \href{https://www.linkedin.com/in/rahulmau/}{Rahul Maurya}, \href{https://www.linkedin.com/in/subhasha-ranjan-7921976}{Subhasha Ranjan}, \href{https://www.linkedin.com/in/daniel-a-levine/}{Daniel Levine}, and \href{https://www.linkedin.com/in/allensiyangwu/}{Allen Wu} for their contributions.
Thank you to the Pulse research team for designing and implementing the \tlag{} metric, building the evaluation infrastructure, and running provider benchmarks end to end.
\bibliographystyle{plainnat}

\end{document}